\documentclass[fleqn,usenatbib,letters]{mnras}

\usepackage{newtxtext,newtxmath}

\usepackage[T1]{fontenc}

\DeclareRobustCommand{\VAN}[3]{#2}
\let\VANthebibliography\thebibliography
\def\thebibliography{\DeclareRobustCommand{\VAN}[3]{##3}\VANthebibliography}

\usepackage{graphicx}	
\usepackage{amsmath}	
\usepackage{CJK}
\usepackage{color}
\usepackage{pdfpages}

\title[Simultaneous emission from dust and gas]{Simultaneous emission from dust and gas in the planetary debris orbiting a white dwarf} 

\author[L. K. Rogers et al.]{Laura K. Rogers,$^{1}$\thanks{E-mail: laura.rogers@ast.cam.ac.uk}
Christopher J. Manser,$^{2,3}$
Amy Bonsor,$^{1}$
Erik Dennihy,$^{4}$
Simon Hodgkin,$^{1}$
\newauthor Markus Kissler-Patig,$^{5}$
Samuel Lai,$^{6}$
Carl Melis,$^{7}$
Siyi Xu \begin{CJK*}{UTF8}{gbsn}(许\CJKfamily{bsmi}偲\CJKfamily{gbsn}艺)\end{CJK*},$^{8}$
Nicola Gentile Fusillo,$^{9}$
\newauthor Boris G\"{a}nsicke,$^{3}$
Andrew Swan,$^{3}$
Odette Toloza,$^{10,11}$
Dimitri Veras$^{3,12,13}$
\\
$^{1}$ Institute of Astronomy, University of Cambridge, Madingley Road, Cambridge CB3 0HA, UK \\
$^{2}$ Astrophysics Group, Department of Physics, Imperial College London, Prince Consort Road, London, SW7 2AZ, UK \\
$^{3}$ Department of Physics, University of Warwick, Coventry CV4 7AL, UK \\
$^{4}$ Rubin Observatory Project Office, 950 N. Cherry Ave., Tucson, AZ 85719, USA \\
$^{5}$ European Space Agency - European Space Astronomy Centre, Camino Bajo del Castillo, s/n., 28692 Villanueva de la Canada, Madrid, Spain \\
$^{6}$ Commonwealth Scientific and Industrial Research Organisation (CSIRO), Space \& Astronomy, P. O. Box 1130, Bentley, WA 6102, Australia \\
$^{7}$ Center for Astrophysics and Space Sciences, University of California, San Diego, CA 92093-0424, USA \\
$^{8}$ Gemini Observatory, 670 N. A'ohoku Place, Hilo, HI 96720, USA \\
$^{9}$ Department of Physics, Universita' degli Studi di Trieste, Via A. Valerio 2, 34127, Trieste, Italy \\
$^{10}$ Departamento de F\'{i}sica, Universidad T\'{e}cnica Federico Santa Mar\'{i}a, Avenida Espa\~{n}a 1680, Valpara\'{i}so, Chile \\
$^{11}$ Millennium Nucleus for Planet Formation, NPF, Valpara\'{i}so, Av. Espa\~{n}a 1680, Chile \\
$^{12}$ Centre for Exoplanets and Habitability, University of Warwick, Coventry CV4 7AL, UK \\
$^{13}$ Centre for Space Domain Awareness, University of Warwick, Coventry CV4 7AL, UK \\
}

\date{Accepted XXX. Received YYY; in original form ZZZ}

\pubyear{2015}

\begin{document}
\label{firstpage}
\pagerange{\pageref{firstpage}--\pageref{lastpage}}
\maketitle

\begin{abstract}

There is increasing evidence for the presence and variability of circumstellar dust and gas around white dwarfs that are polluted with exoplanetary material, although the origin of this dust and gas remains debated. This paper presents the first near-simultaneous observations of both circumstellar dust (via broadband emission) and gas (via emission lines) around a polluted white dwarf. From the optical spectra the gaseous emission lines, notably the calcium infrared triplet and magnesium lines, show significant increases and decreases in their strength over timescales of weeks, while the oxygen and iron lines remain relatively stable. Near-infrared \textit{JHKs} photometry reveals dust emission changes of up to 0.2 magnitudes in the \textit{Ks} band over similar timescales, marking the shortest variability timescales observed to date. The two epochs with the strongest emission were correlated between the dust (\textit{Ks} band brightening) and gas (strengthened calcium and magnesium lines), showing for the first time that the dust and gas must be produced near-simultaneously with a common origin, likely in collisions.

\end{abstract}




\section{Introduction}

There is strong observational evidence for planetary systems around white dwarf stars. Between 25--50 per cent of single white dwarfs are polluted with heavy elements ($>$\,He) resulting from the accretion of planetary material \citep{zuckerman2003metal, zuckerman2010ancient, koester2014frequency, wilson2019unbiased}. 1.5--4 per cent of white dwarfs show an infrared excess from warm ($\sim$\,1000\,K) dust \citep[e.g.][]{becklin2005dusty, kilic2006debris, jura2007externally,rebassa2019infrared, wilson2019unbiased, xu2020infrared,lai2021infrared}. 0.067 per cent of white dwarfs are expected to show Ca\,\textsc{ii} emission lines from a Keplerian rotating gas disc \citep{manser2020frequency}, with 21 identified to date \citep{gaensicke2006gaseous, gansicke2007sdss, gansicke2008sdss, melis2010echoes, farihi2012trio, melis2012gaseous, brinkworth2012spitzer, dennihy2020five,melis2020serendipitous,gentile2020white}, and a handful of white dwarfs show absorption from circumstellar gas \citep[e.g.][]{debes2012detection,xu2016evidence}. X-ray emission has been detected around one polluted white dwarf with a disc, providing direct evidence of ongoing accretion from the disc \citep{Cunningham2022Xrays}. Additionally, a few exoplanets have been discovered around white dwarfs adding to the significant observational evidence of the survival of planetary systems to the white dwarf phase \citep[e.g.][]{gansicke2019accretion,vanderburg2020giant,Blackman2021Jovian}. \textit{JWST} is expected to drastically increase this number \citep{Limbach2022method,Poulsen2024MIRI} with three candidate giant planets already identified \citep{Limbach2024MEOW,Mullally2024JWST}.

The material polluting white dwarfs is thought to be from the scattering of planetesimals onto star-grazing orbits by outer planetary-mass perturbers \citep[e.g.][]{debes2002there}. The planetesimals tidally disrupt at the Roche radius and form an eccentric stream of dust before they circularise, sublimate, and finally accrete onto the white dwarf \cite[e.g.][]{jura2003tidally,brouwers2022road}. The abundances of the accreted material aligns with this scenario as the majority resemble bulk Earth \citep[e.g.][]{jura2014extrasolar}. The origin of the observable circumstellar gas discs is debated. Gas produced at the sublimation radius could viscously spread outwards causing an overlap in the location of the dust and gas \citep{rafikov2011runaway,metzger2012global}. An alternative explanation for gas emission is collisions between large planetesimals that are ground down into small dust within the Roche radius of the white dwarf \citep{jura2008pollution,kenyon2017numericalb,kenyon2017numericala}. Observations of infrared variability in WD\,0145+234 appear consistent with simple collisional cascade models \citep{wang2019ongoing, swan2021collisions,Swan2024first}.

As highlighted above, the circumstellar environments of polluted white dwarfs are known to be variable. Most polluted white dwarfs with infrared emission from a dust disc display variability in the mid-infrared \citep{xu2014drop,swan2019most,swan2020dust,swan2021collisions}, however, few display variability in the near-infrared \citep{rogers2020near}. Additionally, the largest variations in the dust are found for those white dwarfs with discs that have both an observable dusty and gaseous component \citep{swan2020dust,Guidry2024using}. More than half of the currently known white dwarfs with gas discs in emission exhibit variation in the morphology and/or equivalent width of the gas emission lines, attributed to precession or the orbiting of planetesimals within the disc \citep{wilson2014variable,wilson2015composition,manser2015doppler, manser2016another, dennihy2018rapid,manser2019planetesimal,dennihy2020five,gentile2020white,melis2020serendipitous}. Furthermore, observations of transiting debris around a fraction of white dwarfs show variability in the shapes, depths and timescales of the transits which can evolve on periods of days \citep{vanderburg2015disintegrating,Gansicke2016highspeed,vanderbosch2021recurring,Farihi2022relentless}. 

Numerous studies have investigated dust variability and gas variability independently, however, so far these have not been studied simultaneously. This work reports the first study of a metal polluted white dwarf hosting a circumstellar disc of dust and gas, WD\,J210034.65+212256.89 (hereafter called: WD\,J2100+2122, with properties listed in Table\,\ref{tab:Target_Properties}), where the dust and gas emission features have been observed simultaneously. WD\,J2100+2122 was first identified as a white dwarf candidate in \citet{gentile2018gaia}, and a combination of historic photometry and new spectra revealed it had metal pollution, a circumstellar dust disc and a circumstellar gas disc \citep{melis2020serendipitous,dennihy2020five}. The abundance of the material polluting WD\,J2100+2122 was reported in \citet{Rogers2023sevenI} and was found to accrete material similar in composition to bulk-Earth \citep{Rogers2024sevenII}. The gas disc hosts a rich spectrum of emission features showing lines from oxygen, calcium and iron, and a circumstellar dust disc which has an infrared excess above the white dwarf photosphere significant to 7.7\,$\sigma$ in the \textit{K}-band. Additionally, it was demonstrated that the gas disc emission features are variable with the appearance and disappearance of the emission features over timescales of 2 months \citep{dennihy2020five}, and the sparsely sampled \textit{WISE} infrared lightcurves were also found to vary significantly over timescales of a year. Therefore, this was selected as the most ideal target to study simultaneous variability in the dust and gas emission.

\section{Observations and Data Analysis}

\begin{table}
	\centering
	\caption{Properties of WD\,J2100+2122 with astrometry from \textit{Gaia} DR3, and effective temperature and $\log(g)$ from \citet{Rogers2023sevenI}.}
	\label{tab:Target_Properties}
	\begin{tabular}{lc} 
		\hline
        Gaia DR3 number & 1837948790953103232 \\
        RA & 21:00:34.65 \\
        DEC & +21:22:56.89 \\
        Distance (pc) & 88.1 (0.4) \\
        SpT & DAZ \\
        Effective Temperature (K) & 25565 (358) \\
        $\log(g)$ & 8.10 (0.04) \\
		\hline
        Gaia \textit{G} (mag) & 15.172 (0.003) \\ 
        2MASS \textit{J} (mag) & 15.789 (0.066) \\
        2MASS \textit{H} (mag) & 15.545 (0.134) \\
        2MASS \textit{Ks} (mag) & 14.904 (0.119) \\
        \textit{Spitzer} IRAC1 ($\mu$Jy) & 679.2	(34.7) \\
        \textit{Spitzer} IRAC2 ($\mu$Jy) & 685.3	(34.7) \\
        \hline
	\end{tabular}
\end{table}

\subsection{VLT X-shooter Observations}

WD\,J2100+2122 was observed with the echelle spectrograph X-shooter \citep{vernet2011x} on Unit Telescope 3 (UT3) of the Very Large Telescope (VLT) at Paranal Observatory, Chile between October 2019 and September 2023. X-shooter has 3 arms: UVB (3000--5595\,\AA), VIS (5595--10240\,\AA), and NIR (10240--24800\,\AA); the NIR has a low signal-to-noise ratio (SNR) due to the observing set-up being optimised for the UVB and VIS, therefore, the data from this arm are not used. Supplementary Table A1 reports the observations considered in this study. Observations from multiple X-shooter programs with differing observing strategies are used. For the majority of observations STARE mode was implemented, with a 1.0 and 0.9 arcsec slit width for the UVB and VIS arms, respectively, this gives a resolving power ($\lambda / \Delta \lambda $) of 5400 and 8900. These observations consisted of two exposures lasting 571 and 600\,s each for the UVB and the VIS arms, respectively. For the data from programs 0109.2383.005 and 0109.2383.007, a slit width of 1.3 and 1.2 arcsec were used giving a resolving power of 4100 and 6500 in the UVB and VIS arms, with total exposure times of 2946 and 2970\,s each.

The data reduction was performed using \textsc{esoreflex} (v 2.11.3) with the X-shooter pipeline version 2.9.1 \citep{2013A&A...559A..96F} which produced flux calibrated 1D spectra. Supplementary Table A1 reports the SNR for each exposure in the UVB and VIS arms, where the SNR was calculated between the wavelength range 4000 -- 5500\,\AA~for the UVB arm and 6000 to 9000\,\AA~for VIS arm and is $>$\,86 and $>$\,47 per pixel for all exposures in the UVB and VIS arms respectively.

To analyse the emission profiles of the gas, the model spectrum of WD\,J2100+2122 from \cite{Rogers2023sevenI} was subtracted from each X-shooter spectrum. A low order (between 2 and 5 depending on the spectral region) polynomial was fitted to the resultant continuum-subtracted spectrum to remove any deviations from zero flux. Line fluxes were then calculated from these continuum-subtracted spectra by integrating the line flux over the regions where emission is seen as has been done previously \citep[e.g.][]{Xu2024Cloudy}, the line fluxes are reported in Supplementary Table A2. As the line profiles show the characteristic double-peaked line profiles of a rotating disc of gas, the profiles were also converted into velocity space, using the rest wavelengths of the profiles obtained from the NIST database \footnote{\url{https://physics.nist.gov/PhysRefData/ASD/lines_form.html}} and the Doppler shift formula at non-relativistic velocities. These velocities subsequently had their systemic velocity subtracted \citep[$-$31.1\,km/s using values from][]{Rogers2023sevenI} such that the velocities are with respect to the white dwarf. The maximum velocities in the disc, corresponding to the Full-Width Zero Intensity (FWZI) were determined as the point where the profile reaches the continuum.

\begin{figure}
	\includegraphics[width=1.0\columnwidth]{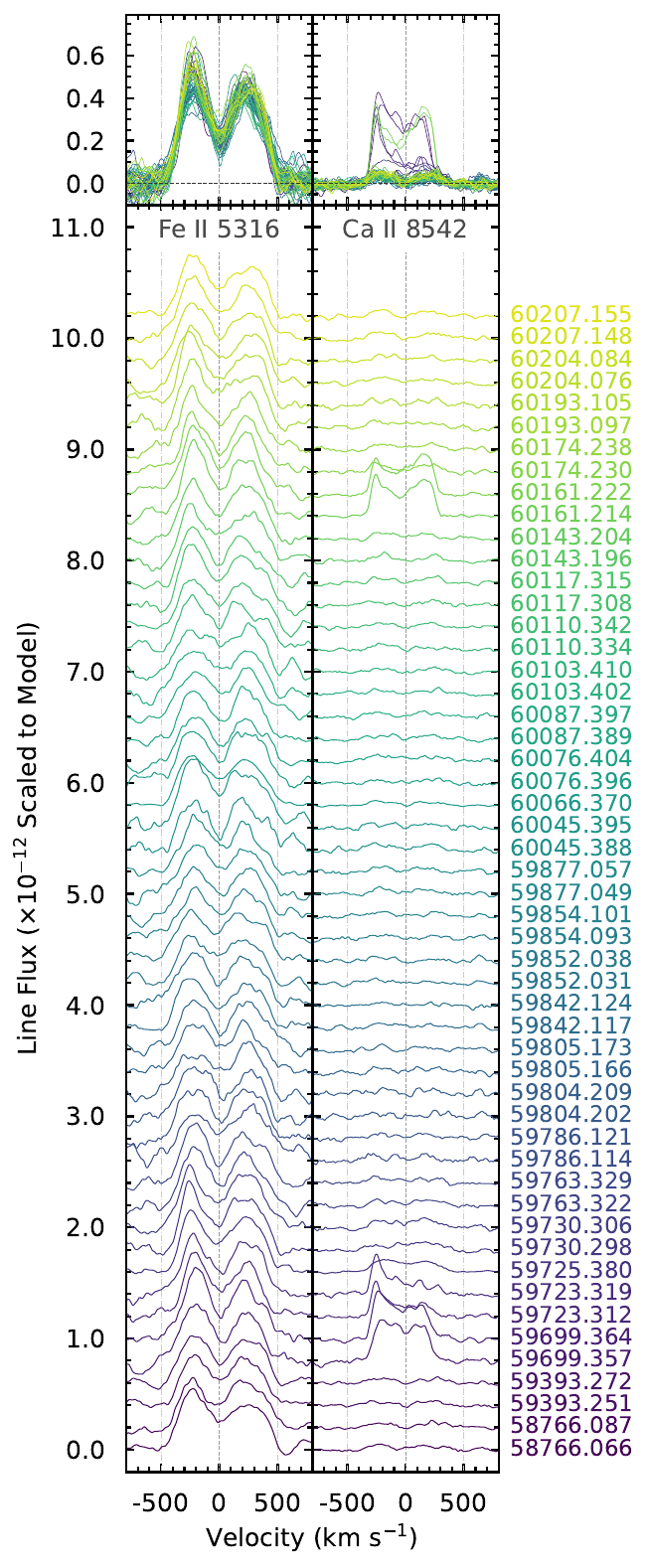}
    \caption{The evolution of the Fe\,\textsc{ii} 5316\,\AA~ and Ca\,\textsc{ii} 8542\,\AA~ lines over time, with each spectrum being offset vertically such that time evolves upwards and the data are labelled with their MJD. As line fluxes are used, the emission strength for each line can be compared directly. There are two regions in time in which the calcium line increases in strength, yet no increase in strength is seen in the iron line. The top panels show all the lines stacked on the same plot, showing clear differences in strengths and FWZIs.
    }
    \label{fig:Gas}
\end{figure}

\subsection{VLT HAWK-I Observations}

WD\,J2100+2122 was also observed with the High Acuity Wide field K-band Imager (HAWK-I) on the Unit Telescope 4 (UT4) of the VLT \citep{Pirard2004HAWKI,Casali2006HAWKI,KisslerPatig2008HAWKI,Siebenmorgen2011HAWKI} between April 2022 and September 2023. Six exposures of 10 seconds were used with five dithers in a jitter box of width 20 arcseconds; this totaled five minutes of integration time per filter. The \textit{J}, \textit{H}, and \textit{Ks} magnitudes were extracted from the phase 3 source tables and differential photometry was performed using stellar objects with \textit{Ks} band magnitudes between 12 and 17 within 700 pixels of WD\,J2100+2122 as comparison stars. This gave 22 sources for differential photometry in which a weighted mean was used to correct the frame-to-frame variations. The magnitude errors were calculated as a sum in quadrature of the measurement error and the systematic error calculated from the comparison stars used for differential photometry. The magnitudes and errors are reported in Supplementary Table A1, with the differential photometry giving errors of up to 0.006, 0.011, and 0.010 mags in the \textit{J}, \textit{H}, and \textit{Ks} respectively. To test the absolute flux calibration, a comparison with 2MASS stellar sources in the field was done. The 2MASS magnitudes were transformed to match the HAWKI filters using equations from the HAWKI user manual \footnote{\url{https://www.eso.org/sci/software/pipelines/hawki/hawki-pipe-recipes.html}}. The standard error for the difference between the transformed 2MASS and HAWKI magnitudes are: \textit{J}: 0.01 mags, \textit{H}: 0.02 mags, \textit{Ks}: 0.02 mags, and these should be added in quadrature with the errors reported in Supplementary Table A1 if absolutely calibrated magnitudes are required. 

\section{Results}
\subsection{Gas variability}

The circumstellar gas disc around WD\,J2100+2122 is found to be highly variable with both emission line strength and morphological changes observed. Figure \ref{fig:Gas} illustrates the evolution of the emission line profiles for the Fe\,\textsc{ii} 5316\,\AA~ and the Ca\,\textsc{ii} 8542\,\AA~ lines over the four year observing period. Notably, the calcium lines exhibit two distinct epochs where the lines increase in strength significantly and decay over times of 64 and 13\,d. The same behaviour is not reflected in the iron lines, which remain at a stable emission strength comparable in flux to the brightest epochs of calcium. The calcium lines show clear asymmetries during the increase in strength on 59723 MJD (and to a lesser degree 59699 MJD), whereas, during the increase in strength on 60161 MJD the calcium lines show a symmetric profile. From the top panel of Fig.\,\ref{fig:Gas}, it is clear that the iron lines have a mostly symmetric profile throughout the observations, with the exception that there appears to be variation in the strength of the blue peak.

The top panel in Fig.\,\ref{fig:Gas} shows a difference in the FWZI for the iron and calcium lines, and this is quantified in Table \ref{tab:FWZI}. Additionally, there is a different FWZI depending on whether the calcium line is in its strong emission state versus quiescent state. Assuming a Keplerian rotating gas disc, the iron lines appear to arise from closer in to the white dwarf in comparison to the calcium lines, and when the calcium is strongly emitting, the gas appears to be originating further out than both the calcium in quiescence and the iron.

\begin{table}
	\centering
	\caption{FWZI measurements for WD\,J2100+2122, where BWZI is the velocity width from the center of the emission line to the blue zero intensity, and RWZI is the same but for the red side, these velocities are related to the true velocity by $\nu _{\mathrm{obs}} = \nu_{\mathrm{true}}\,\mathrm{sin}\,i$. The orbital radii of the inner edge of the gas disc assuming an edge on inclination is also given for reference. The Fe\,\textsc{ii} 5316\,\AA~ line is compared with the Ca\,\textsc{ii} 8542\,\AA~ line for the strongly emitting cases (labelled `E') and the quiescent case (labelled `Q').}
	\label{tab:FWZI}
	\begin{tabular}{cccccc} 
		\hline
        Line & FWZI & BWZI & r$_{\mathrm{b}}$ & RWZI & r$_{\mathrm{r}}$  \\
        \AA & km\,s$^{-1}$ & km\,s$^{-1}$ & R$_{\odot}$ & km\,s$^{-1}$ & R$_{\odot}$ \\
  
        \hline
        5316 & 1020\,$\pm$\,7 & 465\,$\pm$\,5 & 0.61$\,\pm\,$0.01 & 555\,$\pm$\,5 & 0.43\,$\pm$\,0.01 \\
        8542 E & 655\,$\pm$\,7 & 350\,$\pm$\,5 & 1.07\,$\pm$\,0.03 & 305\,$\pm$\,5 & 1.41\,$\pm$\,0.05 \\
        8542 Q & 795\,$\pm$\,7 & 390\,$\pm$\,5 & 0.87\,$\pm$\,0.02 & 405\,$\pm$\,5 & 0.80\,$\pm$\,0.02 \\ 
        \hline
	\end{tabular}
\end{table}

\begin{figure}
	\includegraphics[width=1.0\linewidth]{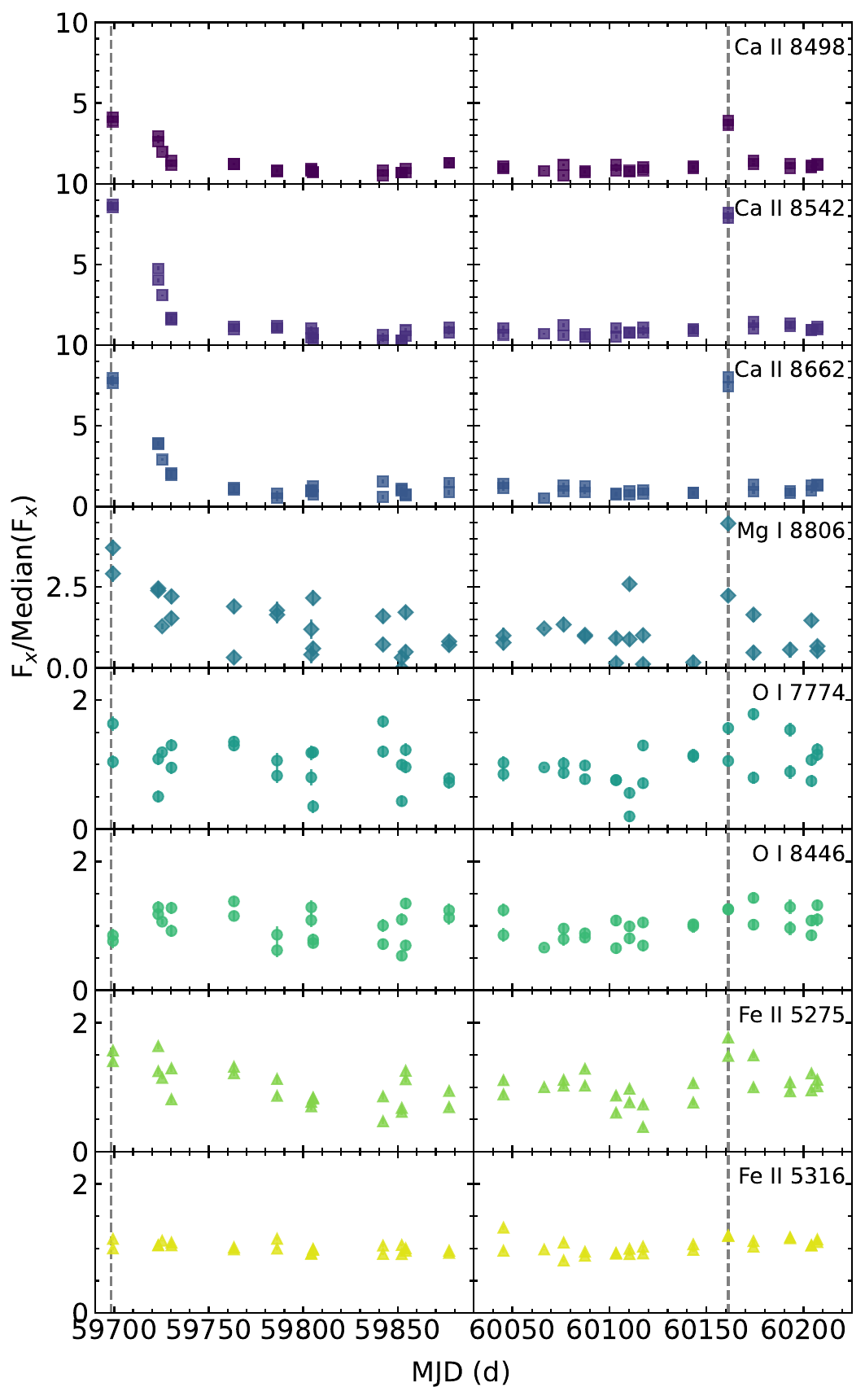}
    \caption{The line flux of the gas emission lines normalised to their median flux over time. The strong emission behaviour of the calcium lines is reflected in all 3 components of the infrared triplet and in the magnesium line, and the two strongest epochs are marked with vertical dashed lines at 59699 and 60161 MJDs. The oxygen and iron lines do not appear to vary in the same way as the calcium lines, but low level variability is clear. Error bars are plotted but are mostly hidden below the data points. }
    \label{fig:Gas_2}
\end{figure}

Figure \ref{fig:Gas_2} presents the line fluxes of the calcium, magnesium, oxygen, and iron lines. All three calcium lines demonstrate similar behaviour with both epochs of increased brightness reflected in all their line fluxes. The magnesium line shows a significant increase in brightness at the same time as the calcium lines, although in quiescence is often undetectable. In contrast, the oxygen and iron lines show some low level variability in their line fluxes, but do not exhibit the same pronounced increases in strength as the calcium lines.

\subsection{Dust variability}

The dust emission from WD\,J2100+2122 shows clear variability in the \textit{J}, \textit{H}, and \textit{Ks} near-infrared bands, with certain epochs showing emission at a $>3\sigma$ significance compared to the baseline flux level. The top panel of Fig.\,\ref{fig:IR_Gas} displays the HAWKI \textit{J}, \textit{H} and \textit{Ks} magnitudes plotted relative to their median values. The dashed vertical lines indicate the two epochs with the strongest emission observed for the Ca\,\textsc{ii} triplet. The \textit{Ks} band exhibits the largest variation with a change of up to 0.2 magnitudes, while the \textit{J} and \textit{H} bands also show variability during these epochs but at a lesser strength. It should be noted that gas emission lines can occur in the near-infrared and may be contributing to a portion of the variability \citep{Owens2023Disk}.

There is tentative evidence for a temperature change associated with the epochs with the most significant near-infrared emission changes; these occur at 59699 and 60161 Modified Julian Dates (MJDs). The ratio of the \textit{Ks}/\textit{H} band fluxes shows a maximum decrease of 27 per cent whilst the \textit{Ks}/\textit{J} shows a maximum decrease of 32 per cent. A blackbody model was fitted to the \textit{J}, \textit{H} and \textit{Ks} band excesses after subtracting off the white dwarf model flux, the median temperature and blackbody radius of the fits to the HAWKI \textit{JHKs} photometry is 1190\,K and 23.5\,R$_{\mathrm{WD}}$ (0.29\,R$_{\odot}$), with the emission peaks at 59699 and 60161 MJDs having a higher best fitting temperature of 1340\,K and 1280\,K respectively. However, the median error on the temperature fits is 166\,K so a temperature increase in the emission peaks is not statistically significant. Further, more precise photometry extending to mid-infrared wavelengths would be required to confirm whether there is a true temperature change of the dust during these emission peaks.

\subsection{Connection between gas and dust emission}

\begin{figure*}
	\includegraphics[width=0.65\linewidth]{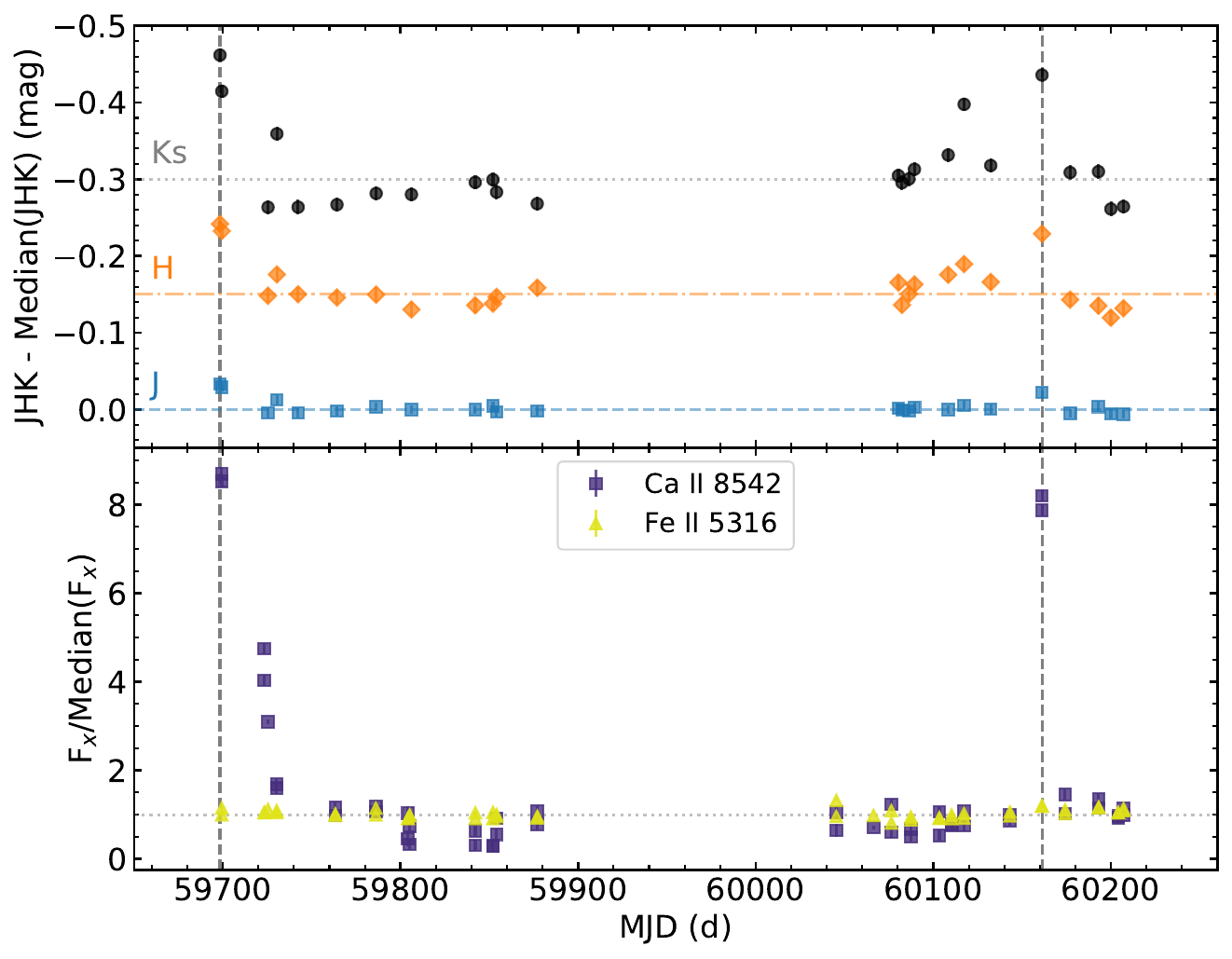}
    \caption{Top: The HAWKI \textit{J}, \textit{H} and \textit{Ks} magnitudes plotted relative to their median magnitude and offset by 0.15 mags for clarity (horizontal lines show the median magnitude per band). Bottom: the gas emission line fluxes of one of the lines from the calcium infrared triplet (8542\,\AA), and one of the iron lines (5316\,\AA) normalised to their median flux over time (with the horizontal dotted line showing the new median flux, which as it is normalised is equal to 1). The rest of the lines are shown in Fig.\,\ref{fig:Gas_2}, and data for line fluxes before this observing period are not shown here as there is no near-simultaneous near-infrared photometry. The dashed vertical lines show where there are clear correlations between the dust emission and the calcium gas emission at 59699 and 60161 MJDs. 
    }
    \label{fig:IR_Gas}
\end{figure*}

Simultaneous increases and decreases in the circumstellar gas emission and infrared flux are found for WD\,J2100+2122. As illustrated in Fig.\,\ref{fig:IR_Gas}, increases in the calcium gas emission coincide with increases in the dust emission, most notable in the \textit{Ks} band (where the infrared excess from the dust is most significant), and also reflected in the \textit{H} and \textit{J} bands. A Pearson's R value of 0.8 demonstrates the strong positive correlation between the gas emission and \textit{Ks} band dust emission. During the two epochs which show concurrent emission, the near-infrared photometry and the optical spectroscopy were obtained on the same observing night and within a few hours of each other emphasising the simultaneity of the dust and gas emission. This correlation highlights for the first time the linked behavior of circumstellar gas and dust emission for any white dwarf.

\section{Discussion}

This study provides for the first time clear evidence for a direct link between dust and gas production in the circumstellar disc surrounding WD\,J2100+2122, with two epochs showing a simultaneous strong increase in emission in both the dust and the (calcium and magnesium) gas and a subsequent return to a base quiescent level. Additionally, it was found that the dust and gas fluxes increase and decrease on timescales of days to weeks, one of the shortest ever recorded for such systems. Furthermore, the behaviour of the gas emission lines is different depending on the element, with the calcium lines showing significant line flux and morphological variations, whereas, the oxygen and iron lines remain relatively stable. 

In this study simultaneous is defined as the two sets of observations taken within a few hours of one another (see Supplementary Table A1 for observing dates) which is sufficient given that the dust and gas are not thought to change on timescales less than the orbital timescale ($\sim$ a few hours). However, as the observations were taken with a cadence of once every two weeks, it remain plausible that there is a delay of $<$\,2 weeks between the increased dust emission and increased gas emission, and only observations with a higher cadence would be able to distinguish this.

The observations of dust and gas trace the inner regions of the circumstellar environment ($<$0.01\,au), however, the origin of the gas surrounding white dwarfs is debated. Following standard theories of white dwarf pollution, the circumstellar dust disc likely formed from a parent body or bodies from a distant surviving planetary system (beyond 3--5\,au) which was perturbed close to the white dwarf and was tidally disrupted which then produced an infrared excess \citep[e.g.][]{debes2002there,jura2003tidally}. However, the gas could be produced by the sublimation of the dust as it migrates inwards and/or by collisions, whether between scattered material and a pre-existing disc or through collisional cascades. There are theoretical predictions for the differences in the dust and gas emission depending on whether the gas is produced by sublimation versus collisions. 

White dwarfs observed with circumstellar gas discs in emission are found to have gas and dust co-located \citep[e.g.][]{melis2010echoes}. \citet{Owens2023Disk} fitted a flat opaque dust ring as prescribed in \citet{jura2003tidally} to the spectral energy distribution of WD\,J2100+2122 finding the inner edge of the dust to be 0.41$^{+0.05}_{-0.07}$\,R$_{\odot}$, the outer edge to be 0.70$^{+0.52}_{-0.11}$\,R$_{\odot}$ (although it should be noted that the outer edge is difficult to fit for due to the lack of longer wavelength data), and the inclination to be 60$^{+15}_{-35}$ degrees. Assuming the same inclination for the gas disc, the inner edge of the gas disc as shown by the stable iron lines (0.46 and 0.32\,R$_{\odot}$ for the blue and red shifted components respectively using values from Table \ref{tab:FWZI}) aligns with the inner edge of the dust disc measured by \citet{Owens2023Disk}. However, the inner edge of the calcium gas in emission lies further out than the dust (1.06 and 0.81\,R$_{\odot}$). If this additional gas emission is produced from the sublimation of the dust, it would likely be caused by the dust migrating towards the white dwarf through Poynting-Robertson (PR) drag and when hot enough, it sublimates, with the gas spreading both inwards and outwards such that it is co-located with the dust \citep{rafikov2011runaway}. This would cause the gas mass to increase a time of $\Delta$T later, associated with the PR drag timescale ($t_{PR} \approx$ 15 days for 0.1$\mu$m dust around WD\,J2100+2122). This is inconsistent with the observations shown here as the additional dust and gas emission during 59699 and 60161 MJDs are correlated and near-simultaneous ($\Delta$T must be less than 2 weeks - shorter than the PR drag timescale), and the inner edge of the calcium gas during the emission peaks occurs further out than the inner edge of the dust. Therefore, it is unlikely that the additional gas emission is produced by sublimation of  dust at the sublimation radius.

Instead, collisions provide a more natural explanation for the simultaneous production of dust and gas seen around WD\,J2100+2122. Material could be ejected from a pre-existing disc by incoming debris, or generated in destructive collisions \citep[e.g.][]{Malamud2021Circularization,swan2021collisions}. Those respective scenarios would result in increased gas emission as dust sublimates above the disc's surface, or as it is collisionally vaporised as has been observed in extreme debris discs around main sequence stars \citep[e.g.][]{Meng2014Large}. Higher cadence data around the emission peaks would be required to assess the dominant timescales for the infrared decay: settling back into an optically thick disc would happen on orbital timescales (hours), whereas collisional cascades have a characteristic 1/t decay \citep{Dominik2003Age}. Collisions have been invoked to explain dust variability \citep{farihi2018dust,swan2021collisions,Swan2024first}, but this work shows for the first time that gas is likely also produced simultaneously in these collisions.

Supporting this scenario, the only white dwarfs demonstrating near-infrared \textit{J}, \textit{H}, or \textit{K} band variability are those that also exhibit gas discs \citep{xu2014drop,xu2018infrared,swan2021collisions}, and \citet{swan2020dust,Guidry2024using} found that those white dwarfs with gas discs also show more variable mid-infrared dust emission, adding to this strong correlation between gas and dust production. Furthermore, \citet{Dennihy2017wired} showed that white dwarfs with gaseous discs either have the largest/brightest and/or the hottest discs, and WD\,J2100+2122 fits with this conclusion with the median temperature and blackbody radius for WD\,J2100+2122 being 1190\,K and 23.5\,R$_{\mathrm{WD}}$ (0.29\,R$_{\odot}$).

The observed variability timescales for the gas and dust are the shortest observed to date for any planetary disc around a white dwarf, with significant \textit{Ks} band changes detected within just one day, and the gas emission shifting between peak emission and quiescent phases within 13 days. These timescales are comparable to the photometric variability observed for white dwarfs with transiting debris which show morphological variations in these transits \citep{Gansicke2016highspeed,vanderbosch2021recurring,Farihi2022relentless}. The two observed epochs of dust and gas emission are also not identical, with the first event (beginning at 59699 MJD) showing a slower decay in the gas over 64 days compared to the 13 day decline following the peak at 60161 MJD (assuming the emission at 59699 MJD is due to a single event). The physical mechanisms responsible for the decay in the dust and gas emission must act on different timescales, highlighting the complexity of the processes at play in these dynamic circumstellar environments. 

The significant variation in the strength of the calcium gas emission for WD\,J2100+2122 would suggest that either (i) a large amount of gaseous material is being generated somewhere in the disc at a temperature to facilitate Ca\,\textsc{ii} triplet emission but not Fe\,\textsc{ii} or O\,\textsc{i} emission, (ii) a region of the disc is heating to enhance Ca\,\textsc{ii} triplet emission, or (iii) cooling to allow Ca\,\textsc{iii} ions to recombine into Ca\,\textsc{ii} (which has an ionisation energy of 11.97 eV). As the dominant source of heating for these planetary debris discs is the photoionising flux of the white dwarf, it is unlikely that options (ii) or (iii) are physical. The upper energy levels for the transitions that results in the observed emission lines are much lower for the calcium lines (3.15, 3.15, and 3.12\,eV for the 8500, 8545 and 8660\,\AA~lines respectively), in comparison to the iron lines (5.55 and 5.48\,eV for 5275 and 5316\,\AA) and the oxygen lines (10.74 and 10.998\,eV for 7774 and 8446\,\AA). Therefore, it is likely that scenario (i) is occurring with the gas disc surrounding WD\,J2100+2122 showing radially dependent excitation with an inner hotter region and an outer more dynamic cooler region. The hotter inner disc is quiescent and gives rise to the broad iron and quiescent calcium line profiles. Given the high temperature of this white dwarf the calcium is likely mostly ionised to Ca\,\textsc{iii} and magnesium ionised to Mg\,\textsc{ii} explaining the weak Ca\,\textsc{ii} and Mg\,\textsc{i} profiles in this hot inner quiescent disc. In the outer, cooler, and more dynamic regions of the disc in which collisions are likely producing gas and dust simultaneously, the right temperature exists to facilitate these strong Ca\,\textsc{ii} lines. In depth modelling of the circumstellar gas using photoionisation codes like the studies by \citet{Hartmann2011nonlte,Hartmann2014ton345,gansicke2019accretion,steele2021characterization,Xu2024Cloudy} will provide further insight into the composition of the gas and the variability of the different species, but is beyond the scope of this work. Additionally, the iron and quiescent calcium line profiles show a more extended red wing, compared to the calcium profile when in its strongly emitting state which has a more extended blue wing. This is likely because the disc is eccentric and non-axisymmetric; this adds to the handful of white dwarfs observed with asymmetric profiles \citep[e.g.][]{melis2010echoes}. Contrasting morphologies and strengths of gaseous emission profiles for different elements has previously been observed, for example, for SDSS\,J1228+1040 \citep[][]{manser2015doppler,Goksu2024gasdisc}. In SDSSJ1228+1040 the calcium, oxygen and iron lines showed a similar spatial extent (FWZI) but differing strengths and morphologies which was explained by each element having a different intensity distribution across the disc.

\section{Conclusions}

This paper reports near simultaneous observations of the gas and dust surrounding the polluted white dwarf WD\,J2100+2122. The key conclusions are as follows: 
\begin{enumerate}
\item There is a positive correlation between gas and dust emission, with the line fluxes of the calcium and magnesium gas increasing with the near-infrared photometry, representing the dust emission. This implies that there is an underlying stable gas and dust disc with additional dust and gas produced simultaneously, likely in collisions. 
\item The gas emission lines produced by the Ca\,\textsc{ii} triplet are extremely variable and change on short timescales of days -- weeks, some of the shortest variability observed in gas emission lines to date. 
\item The variability and behaviour of the gas emission lines depends heavily on the atomic transition observed, with line flux and morphological variations observed in the calcium lines, with limited changes in the iron and oxygen lines. There is likely radially dependent excitation, which could be caused by a hotter inner disc and a colder more dynamic outer disc. 
\item Observing dust and gas simultaneously is a crucial tool to disentangle models of the circumstellar environments around white dwarfs. Further studies with higher cadence around emission peaks are critical to probe the timescales on which the dust and gas evolve. The complexity of the circumstellar environment of WD\,J2100+2122 highlights the stochastic nature of white dwarf planetary systems and paves the way to truly understand how planetary material arrives in the atmospheres of white dwarfs.
\end{enumerate}

\section*{Acknowledgements}

LKR acknowledges support of an ESA Co-Sponsored Research Agreement No. 4000138341/22/NL/GLC/my = Tracing the Geology of Exoplanets and AB and LKR acknowledges support of a Royal Society University Research Fellowship, URF\textbackslash R1\textbackslash 211421. CJM acknowledges the financial support from Imperial College London through an Imperial College Research Fellowship grant. SX is supported by the international Gemini Observatory, a program of NSF NOIRLab, which is managed by the Association of Universities for Research in Astronomy (AURA) under a cooperative agreement with the U.S. National Science Foundation, on behalf of the Gemini partnership of Argentina, Brazil, Canada, Chile, the Republic of Korea, and the United States of America. This project has received funding from the European Research Council (ERC) under the European Union's Horizon 2020 research and innovation programme (Grant agreement No. 101020057). The authors thank Uri Malamud for useful discussions that aided the paper. The authors also thank the referee for their helpful comments that improved the manuscript.

Based on observations collected at the European Southern Observatory under ESO programme IDs: 0103.C-0431, 0104.C-0107, 0109.C-0683(A), 0109.C-0683(B), 0109.2383.005, 0109.2383.007, 0111.C-2308(A), 0111.C-2308(B),

\section*{Data Availability}

VLT X-shooter and VLT HAWK-I data available from the ESO archive (\href{http://archive.eso.org/eso/eso\_archive\_main.html}{http://archive.eso.org/eso/eso\_archive\_main.html}). HAWKI program IDs: 0109.C-0683(B), 0111.C-2308(B) and X-shooter program IDs: 0103.C-0431, 0104.C-0107, 0109.C-0683(A), 0109.2383.005, 0109.2383.007 and 0111.C-2308(A).



\bibliographystyle{mnras}
\bibliography{Master-Bib} 







\bsp	
\label{lastpage}

\clearpage
\includepdf[pages=-]{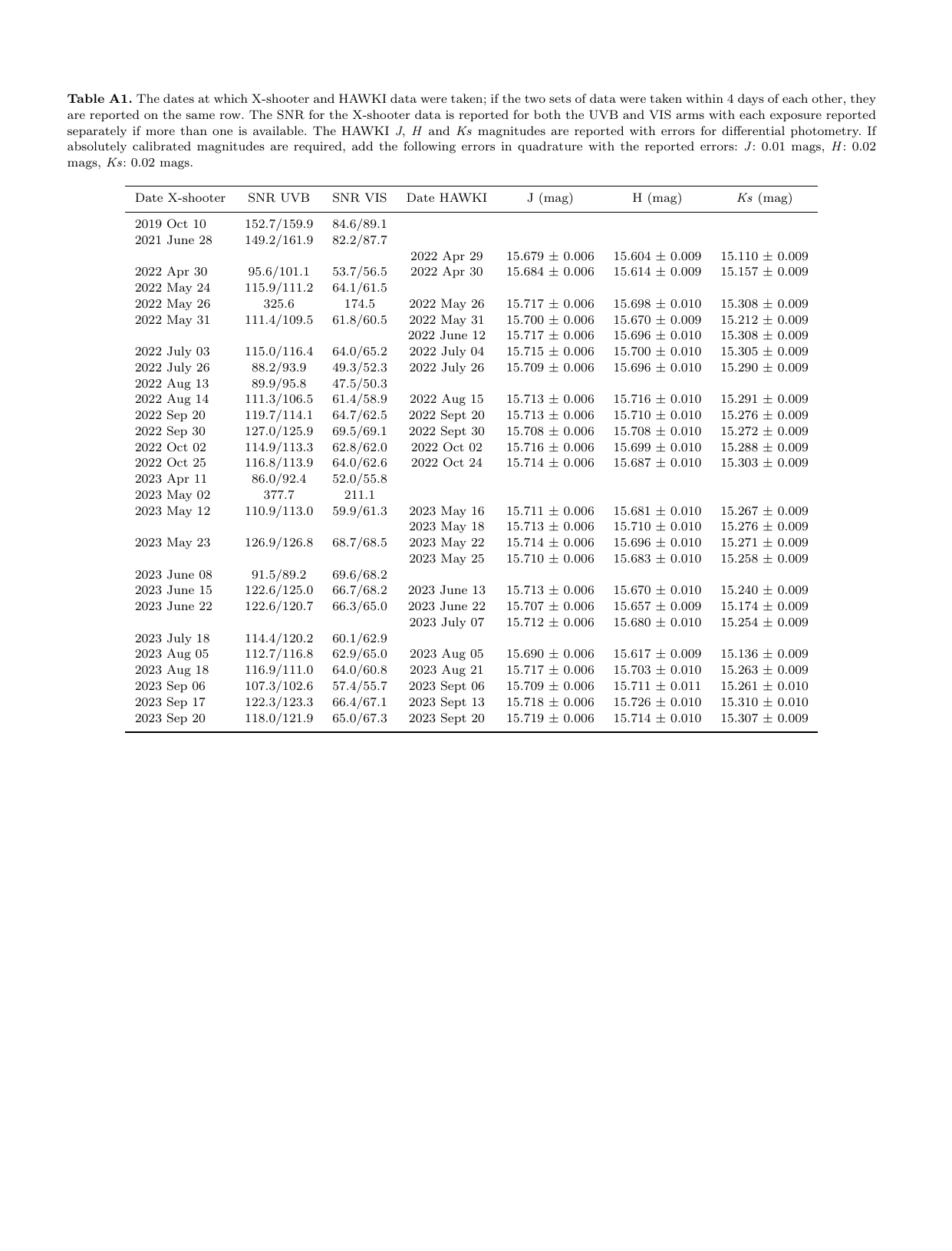}

\end{document}